\documentclass[aps,prl,twocolumn,groupedaddress]{revtex4-2}

\usepackage{graphicx}
\usepackage{amsmath}
\usepackage{hyperref}

\usepackage{xcolor}

\begin{document}

\title{Synchronization in disordered oscillatory media: a nonequilibrium phase transition for driven-dissipative bosons}

\author{John P. Moroney and Paul R. Eastham}

\affiliation{School of Physics, Trinity College Dublin, Dublin 2, Ireland}

\date{\today}

\begin{abstract}
We show that a lattice of phase oscillators with random natural frequencies, described by a generalization of the nearest-neighbor Kuramoto model with an additional cosine coupling term, undergoes a phase transition from a desynchronized to a synchronized state. 
This model may be derived from the complex Ginzburg-Landau equations describing a disordered lattice of driven-dissipative Bose-Einstein condensates of exciton polaritons.
We derive phase diagrams that classify the desynchronized and synchronized states that exist in both one and two dimensions.
This is achieved by outlining the connection of the oscillator model to the quantum description of localization of a particle in a random potential through a mapping to a modified Kardar-Parisi-Zhang equation.
Our results indicate that long-range order in polariton condensates, and other systems of coupled oscillators, is not destroyed by randomness in their natural frequencies.

\end{abstract}


\maketitle


\paragraph{Introduction.---}
Synchronization of coupled oscillators is a phenomenon that appears regularly throughout nature.
Many seemingly unrelated systems which exhibit repetitive behaviors, such as clocks, pacemaker cells in the heart, or a swarm of pulsing fireflies, are seen to undergo transitions from initial randomness to an ordered state \cite{pikovskij_synchronization_2003, strogatz_kuramoto_2000}. A celebrated model of such synchronization phenomena was introduced by Kuramoto \cite{kuramoto_self-entrainment_1975, acebron_kuramoto_2005}. With all-to-all couplings the Kuramoto model undergoes a phase transition, from a desynchronized state at weak coupling to a synchronized state at strong coupling. However, such ordering is absent when the oscillators are on a one or two dimensional lattice, with coupling between neighboring sites\ \cite{sakaguchi_local_1987,strogatz_phase-locking_1988,hong_collective_2005}. 

Synchronization plays an important role in the physics of Bose-Einstein condensation in driven-dissipative Bose gases. Such condensation has been realized for exciton polaritons\ \cite{kasprzak_boseeinstein_2006}, which are bosonic quasiparticles formed from the strong coupling of excitons and photons in semiconductor microcavities. The condensates are described by a single macroscopic wavefunction\ \cite{carusotto_quantum_2013}, giving rise to phenomena such as superfluidity, Josephson oscillations\ \cite{lagoudakis_coherent_2010}, and quantized vortices, and enabling applications such as analog simulation\ \cite{berloff_realizing_2017, amo_exciton-polaritons_2016}. In these systems the decay of the polaritons is offset by gain from the pump, and condensation occurs in a nonequilibrium steady state. Above threshold the nonlinear gain fixes the density of the condensate, forming an autonomous phase oscillator whose frequency corresponds to the condensate's energy. Spatially separated condensates can form in the random potentials arising from disorder in the samples\ \cite{baas_synchronized_2008}, and in engineered potentials such as lattices\ \cite{ohadi_synchronization_2018, berloff_realizing_2017}. If the condensates are well separated they oscillate independently, but their mutual coupling can lead them to synchronize to a common frequency and phase\ \cite{wouters_synchronized_2008}. Such effects have been observed for polaritons in double-well potentials\ \cite{lagoudakis_coherent_2010,wouters_synchronized_2008}, in the intrinsic random potential of the samples\ \cite{baas_synchronized_2008,thunert_cavity_2016}, and in a weakly disordered two dimensional lattice\ \cite{ohadi_synchronization_2018}. 

In this Letter we analyze the coherence properties of driven-dissipative bosons in a disordered lattice potential, focusing on the two dimensional case realized in experiment\ \cite{ohadi_synchronization_2018}. We seek to understand the phase diagram and phase transitions, as generalizations of the equilibrium case, in which there is a superfluid-insulator transition\ \cite{fisher_boson_1989}. The nonequilibrium problem is described by a lattice of coupled oscillators, with disordered frequencies, suggesting that this generalization may be a frequency-ordering transition, or a frequency- and phase-ordering transition. Ordered states are, however, not expected for Kuramoto oscillators with random frequencies in two dimensions\ \cite{sakaguchi_local_1987,hong_collective_2005}. This is consistent with a perturbative treatment of the effect of disorder on the polariton condensate, which predicts the absence of long-range phase order\ \cite{janot_superfluid_2013}. 

Here, we investigate synchronization in a lattice of driven-dissipative condensates with random natural frequencies. Such a lattice is described by a generalization of the nearest neighbor Kuramoto model that includes an additional coupling term that is even in the relative phases.  Such models have recently been considered\ \cite{he_space-time_2017,lauter_kardar-parisi-zhang_2017} in the case where the natural frequencies are identical and frequency synchronization is expected, but the phases may become disordered due to the space- and time-dependent noise associated with gain and loss\ \cite{gladilin_spatial_2014, he_scaling_2015, altman_two-dimensional_2015, squizzato_kardar-parisi-zhang_2018}. We consider the opposite limit, of random natural frequencies but negligible time-dependent noise. We find that the non-Kuramoto coupling has a significant effect: it leads to a true synchronization transition to a state with long-range frequency and phase order for dimensions $d<4$. Our argument involves relating the oscillator model to a Kardar-Parisi-Zhang (KPZ) equation\ \cite{kardar_dynamic_1986} with time-independent noise\ \cite{ebeling_diffusion_1984,nattermann_diffusion_1989,krug_directed_1993,halpin-healy_kinetic_1995,cates_statistics_1988}, and then to an imaginary-time Schr\"odinger equation with a random potential. This allows us to derive the phase boundary for synchronization and characterize the frequency and phase profiles. These analytical predictions agree with numerical simulations in one and two dimensions. Our results show that there is a nonequilibrium generalization of the superfluid-insulator transition in a lattice of driven-dissipative condensates, and that there is a region where there is long-range  coherence that is robust against disorder. Our conclusions apply more generally to coupled oscillator systems, implying there are many other settings\ \cite{pikovskij_synchronization_2003} in which this phase transition could be realized.

\paragraph{Dynamics of nonequilibrium condensates.---}
Under the requisite conditions \cite{bobrovska_adiabatic_2015}, a polariton condensate may be described by a driven-dissipative Gross-Pitaevskii equation \cite{keeling_spontaneous_2008},
\begin{align} 
i\frac{\partial \Psi}{\partial t} =  \bigg[-\frac{1}{2m}\nabla^2 + V_0(\mathbf{r})\bigg]\Psi & + U_0|\Psi|^2\Psi \nonumber \\ & + i \left(g_0(\mathbf{r})-\Gamma_0|\Psi|^2\right)\Psi,\label{eq:gpe}
\end{align} 
where $U_0$ is the polariton-polariton interaction strength, while $g_0(\mathbf{r})$ and $\Gamma_0$ account for the linear gain -- resulting from the difference of pumping and decay -- and the gain saturation.
$V_0(\mathbf{r})$ is a confining potential, which can arise from the repulsive interaction with the exciton reservoir, etching, deposition, and the intrinsic disorder in the sample\ \cite{galbiati_polariton_2012, ohadi_synchronization_2018}. We set $\hbar=1$ throughout.

We consider a lattice of $N$ condensates, as realized experimentally\ \cite{ohadi_synchronization_2018}. In that experiment a spatially patterned pump beam populates the exciton reservoir, forming a confining lattice potential. This potential is superimposed on the random potential due to disorder in the sample, leading to small variations in the energy from site to site. Condensation occurs in the lowest energy state of each well of the lattice potential. We model this by expanding
the macroscopic wavefunction over the basis set of wavefunctions localized in individual wells\ \cite{eastham_disorder_2017,suppref}.
Assuming that the overlap between neighboring wavefunctions is small results in $N$ coupled equations for the amplitudes of each site, $\psi_k$, 
\begin{eqnarray} 
i\frac{\partial \psi_k}{\partial t} &=& \left[\epsilon_k + U|\psi_k|^2 + i \nonumber \left(g-\Gamma|\psi_k|^2\right)\right]\psi_k \\ && - \sum_{<l>}J_{kl}\psi_l.\label{eq:1d}
\end{eqnarray}
Here, $\epsilon_k$ is the energy of the condensate on a given lattice site, $U$, $g$ and $\Gamma$ are the interaction strength and gain coefficients, and the sum is over nearest neighbors. $J_{kl}\approx J$ is the matrix element describing tunneling between neighboring sites. 
Since the pump pattern, which produces the largest contribution to the potential and controls the gain on each site, is periodic, these parameters will vary little from site to site, and we treat them as constant. The site energies however vary randomly from site to site, with their standard deviation $\sigma$ providing a measure of the strength of the disorder.

To examine synchronization in this system, we reparameterize our equations in terms of  density and phase , $\psi_k = \sqrt{n_k}\exp{(-i\theta_k)}$. Well above threshold, the density variation between condensates is small, $\delta n_{kl}=n_{k}-n_{l}\ll n_k$. Furthermore, the fast relaxation of the densities allows for their adiabatic elimination.
This gives\ \cite{suppref}
\begin{equation} \label{eq:pm}
\frac{\partial\theta_k}{\partial t} = \epsilon_k + \frac{g}{\alpha} + J\sum_{<l>}\left[\frac{1}{\alpha}\sin(\theta_l-\theta_k)-\cos(\theta_l-\theta_k)\right].
\end{equation} Here, we have introduced the dimensionless parameter $\alpha \equiv \Gamma/U$. The overall blueshift $g/\alpha$ can be removed by a redefinition of the zero of frequency. We can choose $1/\sigma$ as our unit of time, so that the solutions to Eq. (\ref{eq:pm}) are controlled by two dimensionless parameters, $\alpha$ and $J/\sigma$.

Eq. (\ref{eq:pm}) is an equation for the phase dynamics in a system of coupled self sustained oscillators.
If we neglect the cosine term in the sum, it is the nearest-neighbor Kuramoto model \cite{sakaguchi_local_1987}. In general, we expect that the existence of a synchronized solution -- whereby the oscillators rotate at a common frequency, $\dot{\theta}_1 = \dot{\theta}_2 = \dots = \dot{\theta}_N = \Omega$ -- will be dependent on the magnitudes of the tunneling and the spread of on-site energies. If there is no tunneling ($J=0)$, each oscillator will rotate at its blueshifted natural frequency.
When $J$ is large relative to the spread of the natural frequencies however, the frequency of each oscillator is strongly affected by the phase difference between it and its neighbors, and this mechanism can bring about synchronization. Despite this, it has been shown that one and two dimensional lattices of Kuramoto oscillators, with random on-site energies and nearest-neighbor couplings, do not exhibit synchronization\ \cite{strogatz_phase-locking_1988,hong_collective_2005,sakaguchi_local_1987} in the limit $N\to\infty$. In one dimension the probability of synchronization can be calculated as a function of $N$, and a coupling strength of order $\sqrt{N}$ is required for a synchronized solution\ \cite{strogatz_phase-locking_1988}.
\begin{figure}
\includegraphics[width=\columnwidth]{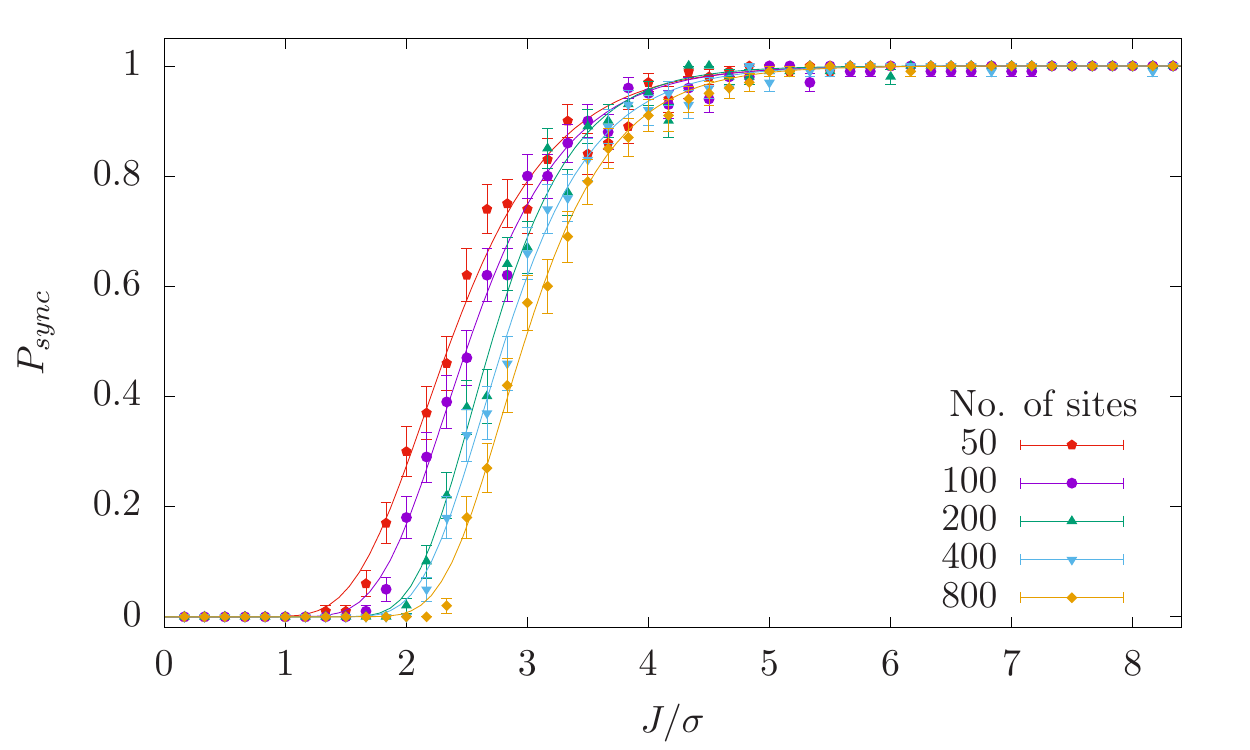}
\caption{Probability of synchronization of chains of condensates of varying lengths, determined by solving Eq. (\ref{eq:pm}) with $\alpha=1$. Each probability is estimated from 100 disorder realizations, each simulated up to a time $t\sigma=1.8\times 10^4$. The curves are fits to the Gumbel distribution. \label{fig:probsync}}
\end{figure}

The presence of the cosine term in the coupling function of our model
has a significant effect however.  Unlike the Kuramoto model, our
coupling is non-odd in its arguments, and it has been suggested that
this may bring about synchronization more readily\
\cite{kopell_symmetry_1986, ostborn_frequency_2004}.  Numerical
simulations of Eq. (\ref{eq:pm}) illustrate that this is indeed the
case.  Fig. \ref{fig:probsync} shows the probability of
synchronization for chains of condensates of various lengths with
normally-distributed on-site energies.  This probability is
determined by calculating the average frequency of each oscillator
from its phases at two times, one a short time after the initial
transient behavior has decayed, and the other a long time later.  If
each bond between neighboring sites has a frequency difference less
than the smallest numerically resolvable frequency the configuration
is considered synchronized.  $P_{sync}(J)$ is then estimated from the
fraction of synchronized results arising over many disorder
realizations.  We see that the probability of synchronization varies from
  almost zero to almost one over a range of $J$. The width of this
  range is non-zero, because different realizations synchronize at
  slightly different tunneling strengths.  More importantly, the
  position of the center of this range, which we use to define a typical
  critical tunnneling strength $J_c$ such that $P_{sync}(J_c) = 0.5$,
  is almost independent of the system size. This strongly suggests
  that there is a synchronization transition in the limit of
  thermodynamically large systems. This would be a non-equilibrium
  phase transition, at which the steady-state of a system in the
  thermodynamic limit changes character, giving rise to singularities
  in its properties.  Such behavior is markedly different from that
of the Kuramoto model, where $J_c$ scales with $\sqrt{N}$, so
  that synchronization occurs only in small systems.

\begin{figure}[t]
    \centering
    \includegraphics[width=\columnwidth]{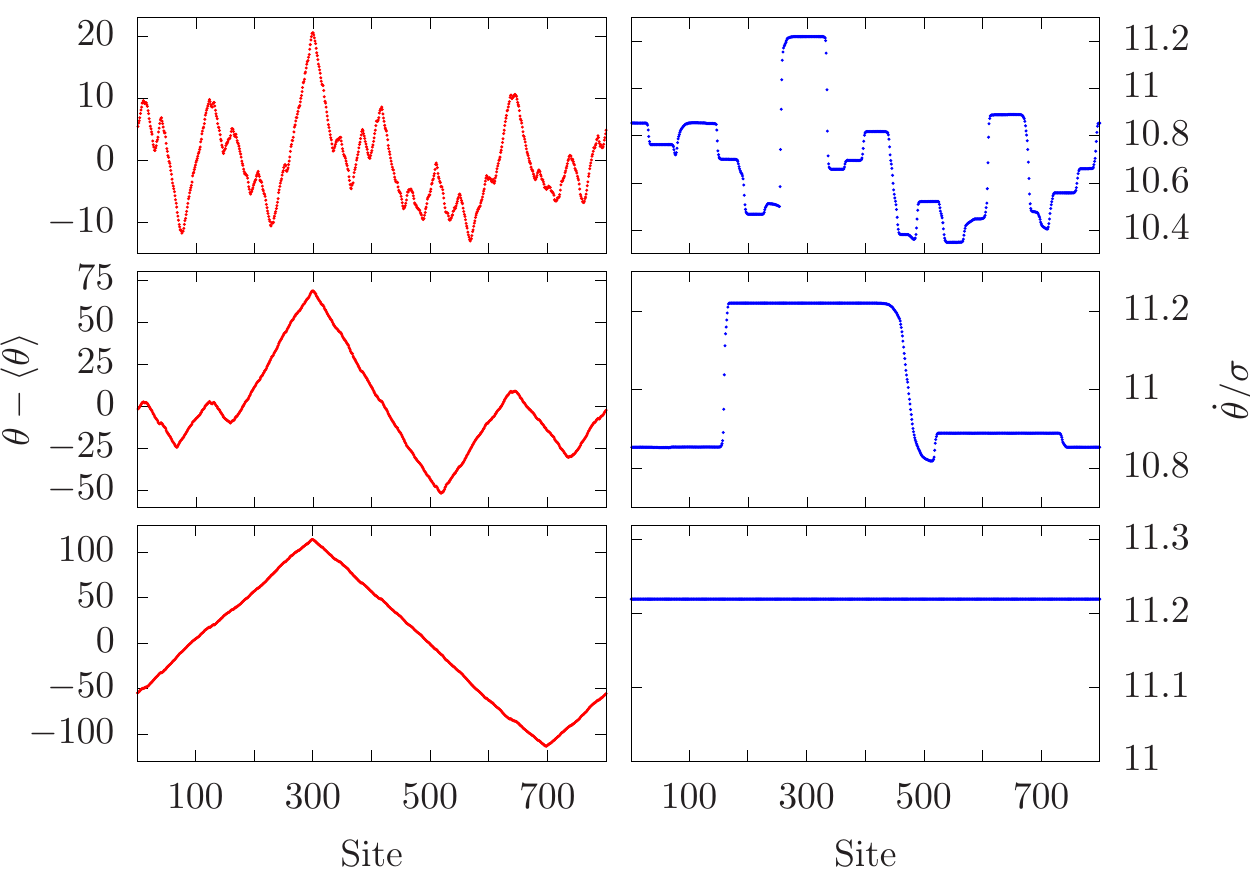}
    \caption{Phase (left column) and frequency (right column) profiles in a chain of 800 coupled oscillators, with $J/\sigma=3.33$ and $\alpha=1$. The oscillators initially have a uniform phase. The profiles are shown after times $t\sigma=100$ (top row), $600$ (middle row), and $24000$ (bottom row).}
    \label{fig:phaseprofs}
\end{figure}

\paragraph{Continuum theory---}
To understand the synchronization transition, and the nature of the synchronized states, we consider the continuum form of Eq. (\ref{eq:pm}) in the limit where the phase differences between neighboring sites are small. 
This will be the case in a synchronized state for weak disorder.
Expanding the trigonometric functions to second order, and taking the continuum limit, we have 
\begin{equation} \label{eq:cont}
    \frac{\partial\theta(\mathbf{x},t)}{\partial t}=\epsilon(\mathbf{x})+\frac{Ja^2}{\alpha} \nabla^2\theta+Ja^2\left(\nabla\theta\right)^2,
\end{equation} 
where $a$ is the lattice constant, which will be set to one in the following.
A uniform energy shift $g/\alpha - 2Jd$ has been absorbed in the definition of $\theta(\mathbf{x},t)$, $d$ being the dimensionality of the system -- $1$ or $2$ for the cases of interest to us. 

Eq. (\ref{eq:cont}) is similar to the KPZ equation, but has a time independent noise term\ \cite{kardar_dynamic_1986, halpin-healy_kinetic_1995}.
The connection between the complex Ginzburg-Landau equation and the conventional KPZ equation, with spatiotemporal noise, has been made in previous studies of polariton condensation\ \cite{gladilin_spatial_2014, he_scaling_2015, altman_two-dimensional_2015, squizzato_kardar-parisi-zhang_2018}, and indeed lattice models similar to Eq. (\ref{eq:pm}) have also been studied \cite{he_space-time_2017, sieberer_lattice_2016, lauter_kardar-parisi-zhang_2017}. 
Those works consider noise associated with gain and loss, rather than that due to a random potential. The form of the noise in Eq. (\ref{eq:cont}) results from our consideration of purely spatial disorder, and  leads to different universal behavior.
The phase, $\theta(x, t)$, behaves like the height of an interface, with a growth rate $\epsilon(\mathbf{x})$ which is random in space but not in time.

\begin{figure}
    \centering
    \includegraphics[trim=0cm 5.5cm 0cm 0cm]{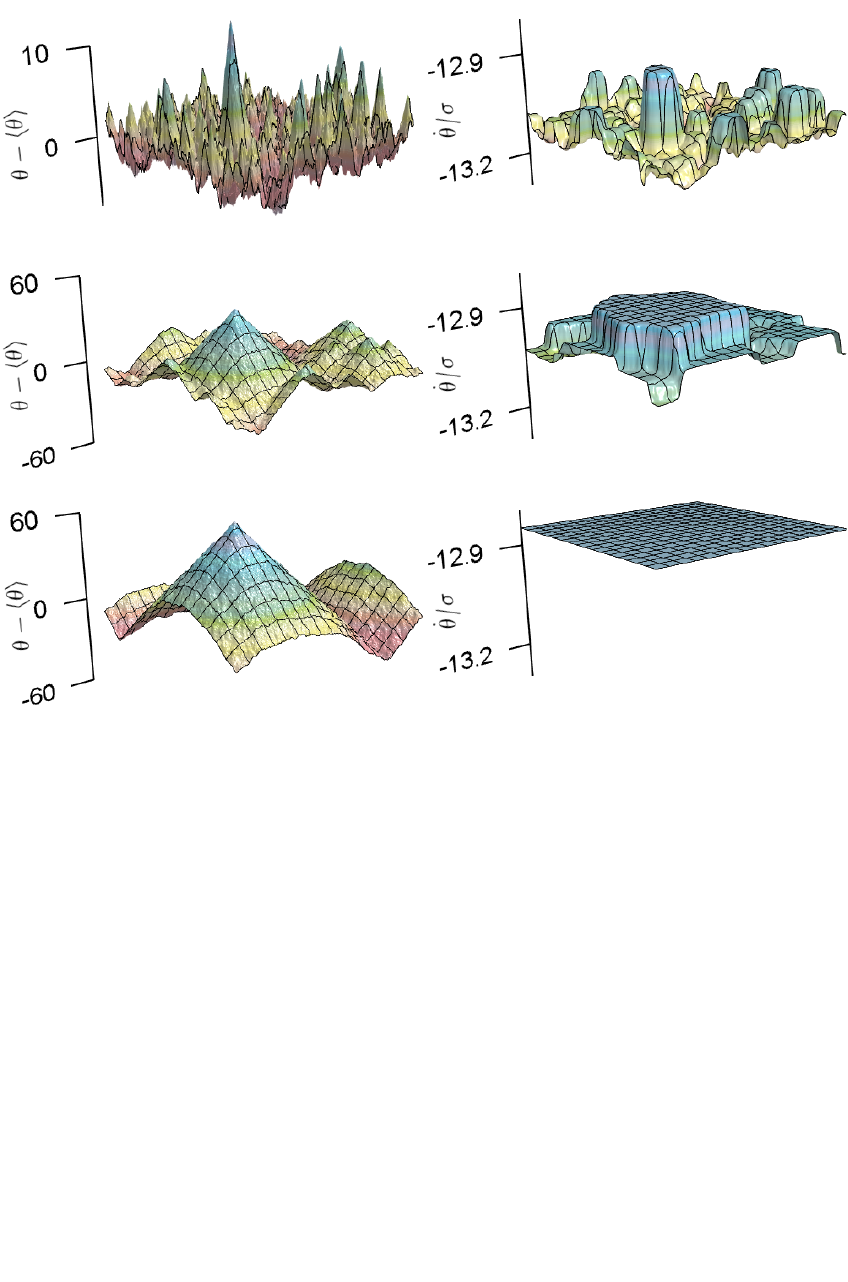}
    \caption{Phase (left column) and frequency (right column) profiles in a two dimensional lattice of $512\times512$ coupled oscillators, with $J/\sigma=3.33$ and $\alpha=1$. The oscillators initially have a uniform phase. The profiles are shown after times $t\sigma=50$ (top row), $300$ (middle row), and $1700$ (bottom row).}
    \label{fig:phaseprofs2d}
\end{figure}

A Cole-Hopf transformation 
\begin{equation} \label{colehopf}
    Z(\mathbf{x}, t)=\exp{(\alpha \theta)},
\end{equation}
enables us to write the KPZ equation as 
\begin{equation} 
    \frac{\partial Z(\mathbf{x}, t)}{\partial t}=\frac{J}{\alpha}\nabla^2 Z + \alpha\epsilon(\mathbf{x})Z=-\hat{H}Z. \label{eq:imschr}
\end{equation}
This is the imaginary time Schr\"odinger equation
for a particle of mass $\alpha/2J$ in a random potential $V(\mathbf{x})=-\alpha \epsilon(\mathbf{x})$. 
In this form it also describes the evolution of a population with diffusion and random autocatalytic amplification \cite{ebeling_diffusion_1984, nattermann_diffusion_1989}, and the partition function for a directed polymer in a random potential \cite{krug_directed_1993, cates_statistics_1988}.
The general solution of Eq. (\ref{eq:imschr}) can be expressed in terms of the eigenstates of $\hat{H}$, with energies $E_n$,
\begin{equation} \label{eq:sol}
    Z(\mathbf{x},t) = \sum_{n}c_n\psi_n(\mathbf{x})e^{-E_nt},
\end{equation} which approaches the ground state wavefunction of the random potential $V(\mathbf{x})$ for $t\rightarrow \infty$. 
For long but finite times, $Z(\mathbf{x},t)$ will have contributions from a small number of low-energy states. 
In dimensions $d<4$ these low-lying states will be localized, $\psi_n \sim e^{-|\mathbf{x}-\mathbf{x}_n|/\zeta}$, at some dilute positions $\mathbf{x}_n$, with the localization length $\zeta$ given by balancing the kinetic and potential terms in Eq. (\ref{eq:imschr}):  $D/\zeta^2\sim u_L/\zeta^{d/2}$\ \cite{nattermann_diffusion_1989}. Here $D=J/\alpha$ and $u_L=\alpha \sigma$, so we have 
\begin{equation} \label{eq:loclen}
\zeta\sim \left(\frac{D}{u_L}\right)^{2/(4-d)}= \left( \frac{J}{\alpha^2 \sigma} \right)^{2/(4-d)}.
\end{equation}
In principle the ground state localization length should be obtained from the depth of the deepest of the $\sim (L/\zeta)^d$ potential wells of size $\sim \zeta$ in the sample, rather than the depth of a typical well. 
These differ by a factor $\sqrt{d \ln{L/\zeta}}$ which, while it formally diverges as $N=L^d\to\infty$, is of order one in realistic systems\ \cite{nattermann_diffusion_1989}.

In Figs. \ref{fig:phaseprofs} and \ref{fig:phaseprofs2d} we show some phase and frequency profiles obtained by solving the coupled oscillator model, Eq. (\ref{eq:pm}). These solutions agree qualitatively with those of the continuum model, given by Eqs. (\ref{colehopf}) and (\ref{eq:sol}). 
At short times, multiple peaks are evident in the phase profile, each of which corresponds to a localized, low-energy state of the Hamiltonian. The corresponding energies appear as plateaus in the frequency profile.
The low-lying states have negative energies, and so they grow in magnitude, corresponding to a steadily increasing phase in the region controlled by each localization center. The ground state, with the lowest energy, grows fastest, and eventually dominates the solution, giving a solution with a single peak in the phase profile, and a single frequency (corresponding to the ground state energy of $\hat{H}$). 
\paragraph{Phase diagram---} The solutions to the continuum model always synchronize, given sufficient time, but this is not the case for the original model. To investigate the conditions for synchronization we consider the phase gradients. Since the solutions to Eq. (\ref{eq:imschr}) are formed from exponentially localized states with localization length $\zeta$  we have, from Eqs. (\ref{colehopf}) and (\ref{eq:loclen}), 
\begin{equation}
    |\nabla\theta| \sim \frac{1}{\alpha \zeta} \sim  \left[\alpha^d\left(\frac{\sigma}{J}\right)^2\right]^{1/(4-d)}.
\end{equation} The continuum model does not impose a limit on these gradients, however, it only captures the behavior of the original model when $|\nabla\theta|\lesssim 1$. Otherwise the periodicity and limited range of the trigonometric functions are important, and we cannot expand them in a power series. This argument gives 
\begin{equation}
    J_c \sim \sigma \alpha^{d/2} \label{eq:phasebd}
\end{equation} as the phase boundary for synchronization.  This condition is relevant for $d < 4$, as localized solutions to Eq. (\ref{eq:imschr}) are not guaranteed otherwise. 
\begin{figure}
    \centering
    \includegraphics[width=0.5\textwidth]{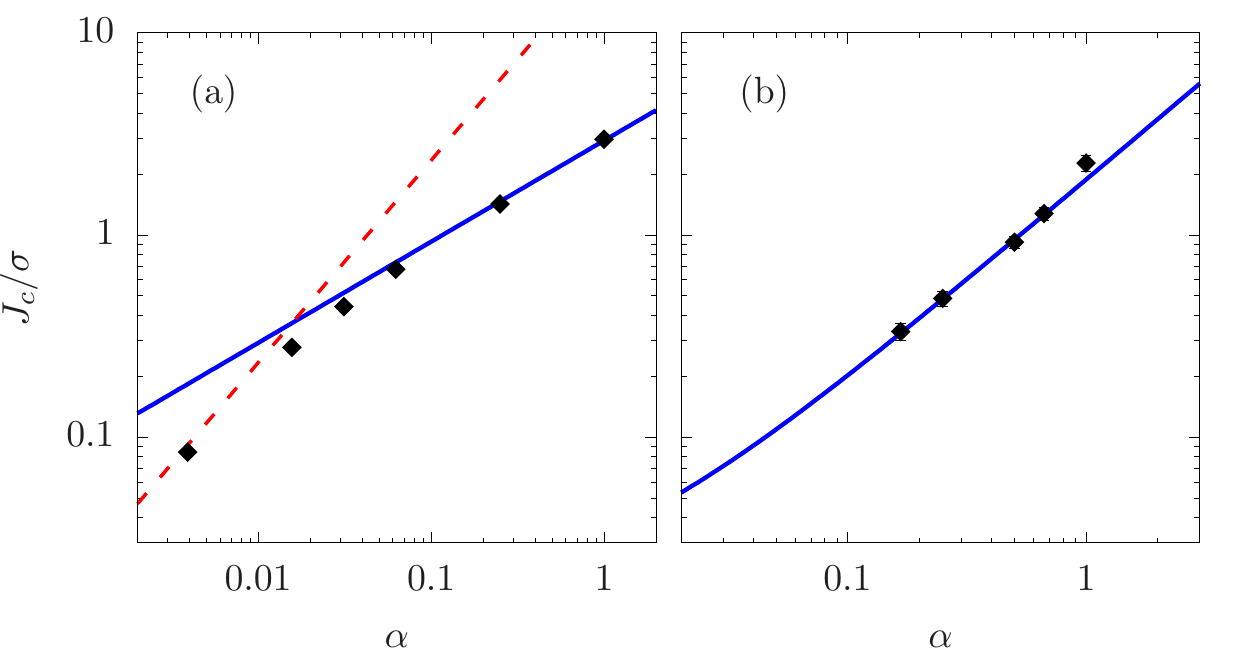}
    \caption{Phase boundary for synchronization for (a) a 1D chain of 800 oscillators, and (b) a 2D square lattice of $128\times128$ oscillators. $J_c$ is defined as the value of $J$ corresponding to a probability of synchronization of 0.5. The points are numerical values, computed as discussed in the text. The solid lines show fits to the predicted slopes of 0.5 (a) or 1 (b). The dashed line in (a) is the phase boundary for synchronization for a chain of 800 Kuramoto oscillators.}
    \label{fig:phasediags2D}
\end{figure}

The dependence of $J_c$ on $\alpha$ given by Eq. (\ref{eq:phasebd}) is
confirmed by simulations for lattices in one and two dimensions, as
shown in Fig. \ref{fig:phasediags2D}. We simulate Eq. (\ref{eq:pm}) to
obtain $P_{sync}(J)$, fit the resulting functions to a Gumbel extreme-value
distribution, and use the fitted parameters to find $J_c$. This extreme-value distribution is expected
  from the analysis above, since the system is synchronized at a given $J$ if the ground-state localization length exceeds a certain value, or equivalently if the ground-state energy exceeds a certain value. It fits our
  results well, as can be seen in Fig.\ \ref{fig:probsync}. The
points in Fig.\ \ref{fig:phasediags2D} show the values of $J_c$. The solid curves show the predicted
square-root (1D) and linear (2D) dependencies, which are in good
agreement with the data.  On the 1D phase diagram, we also show the
prediction of the conventional Kuramoto model for a chain of $N=800$
sites, for which 
$J_c\sim \sigma\alpha \sqrt{N}$. While the critical coupling in that
case, without the cosine term, diverges as $N \rightarrow \infty$, for
a finite system it can nonetheless lie below Eq. (\ref{eq:phasebd}) at
small $\alpha$. This explains the crossover seen at small $\alpha$ in
Fig.\ \ref{fig:phasediags2D}a. This is the standard finite-size
behavior if the cosine term is relevant in the renormalization group
sense: such terms, even if they are small at the lattice scale, grow
with distance, and control the physics in the limit $N\to\infty$.  In
a finite system, however, the growth can be cut off by the system
size.

Although our numerical results agree with Eq. (\ref{eq:phasebd}) for $\alpha \lesssim 1$, they disagree in the opposite regime, where we find large sample-to-sample fluctuations and many states which are desynchronized even at large $J$. This may be related to dynamical instabilities in that regime\ \cite{he_space-time_2017,lauter_kardar-parisi-zhang_2017}. In any case, Eq. (\ref{eq:loclen}) holds only in
the weak disorder regime where the localization length $\zeta$ is larger than the lattice spacing $a=1$, and the localization length at the transition is, from Eqs. (\ref{eq:loclen}) and (\ref{eq:phasebd}), $\zeta_c\sim 1/\alpha$. 
More generally, $\zeta_c>1$ is needed so that space can be treated as continuous, as in Eq. (\ref{eq:cont}), at the phase boundary. 
Importantly, the existence of a synchronization transition in the continuum regime implies the phenomenon is universal, and independent of the details of the lattice or disorder.
We note that a synchronization transition for one dimensional chains of oscillators with non-odd nearest-neighbor coupling was previously identified by \"{O}stborn \cite{ostborn_frequency_2004}. That analysis, however, predicts a critical $J_{c}$ which differs from Eq. (\ref{eq:phasebd}), and does not agree well with our numerical results.

\paragraph{Conclusions---}

We have shown that a model for coupled phase oscillators, which describes a disordered lattice of polariton condensates, has a synchronization transition in the thermodynamic limit. At this transition tunneling between condensates overcomes the localizing effects of the random potential, leading to a state with long-range coherence. This is reminiscent of the Bose-glass--superfluid transition, but here occurs in a driven-dissipative system. The existence of a state with long-range coherence, which is robust against disorder, may be important for applications of polariton condensation in areas such as analog simulation.

The synchronization transition we predict is absent for the Kuramoto model: it arises from the non-odd coupling between the phases, which gives a relevant nonlinear term in the continuum limit. That same form will appear for any system of coupled oscillators, $\dot{\theta_k} = \omega_k + K\sum_{<l>}f(\theta_l-\theta_k)$, in which the coupling function, $f(x)$, is neither purely even nor odd. That will generally be the case, so that our work implies that many other coupled oscillator systems can support synchronized states, notwithstanding disorder in their frequencies.

\begin{acknowledgments}
We acknowledge funding from the Irish Research Council under award GOIPG/2019/2824. Some calculations were performed on the Lonsdale cluster maintained by the Trinity Centre for High Performance Computing. This cluster was funded by grants from Science Foundation Ireland.
\end{acknowledgments}

%

\end{document}



\title{Supplemental Material for ``Synchronization in disordered oscillatory media: a nonequilibrium phase transition for driven-dissipative bosons''}


\author{John P. Moroney and Paul R. Eastham}
\affiliation{School of Physics, Trinity College Dublin, Dublin 2, Ireland}


\date{\today}



\maketitle

\renewcommand{\theequation}{S\arabic{equation}}
\renewcommand{\thefigure}{S\arabic{figure}}
\renewcommand{\bibnumfmt}[1]{[S#1]}


\section{Derivation of Lattice Model}

We obtain the lattice model, Eq. (2) in the main text, from the driven-dissipative Gross-Pitaevskii equation, Eq. (1), by expressing the wavefunction as a superposition of wavefunctions corresponding to the ground state of each well, 
\begin{equation} \label{eq:expand}
\Psi(\mathbf{r}, t) = \sum_k^N\psi_k(t)\phi_k(\mathbf{r}),
\end{equation} where the $\phi_k(\mathbf{r})$ are normalized. We substitute (\ref{eq:expand}) into the Gross-Pitaevskii equation, multiply on the 
left by $\phi^*_l(\mathbf{r})$, and integrate over the sample [34]. 
The wavefunctions $\phi_k$ are localized on each site, so that terms involving their overlap on different sites are small and can be neglected. This leads to Eq. (2) of the main text. The energies in the lattice model,
$\epsilon_k$ are those for the single well. The tunneling term is $J_{kl} = -\int\phi^*_k(\mathbf{r})\left[-(1/2m)\nabla^2+V_0(\mathbf{r})\right]\phi_l(\mathbf{r})d\mathbf{r}$, and is retained only for nearest-neighbor sites since it decays rapidly with distance. The linear gain on each site is $g_k = \int|\phi_k(\mathbf{r})|^2g_0(\mathbf{r}) d\mathbf{r}$, and the interaction is $U_k = \int|\phi_k(\mathbf{r})|^4U_0d\mathbf{r}$, with a corresponding expression for $\Gamma_k$. 

As discussed in the main text, the potential we consider consists of a large periodic component, which is the same for each site, with the addition of a smaller random component, which varies from site to site. Since the former is much larger than the latter, the wavefunctions $\phi_k(\mathbf{r})$ will vary little from site to site, and we treat them as identical. Given also that the pump profile is the same for each site -- $g_0(\mathbf{r})$ has the periodicity of the lattice -- then the parameters of the lattice model $g_k,U_k,\Gamma_k$ and $J_{kl}$ are the same for every site. 
\section{Phase-only model}
As discussed in the main text, we obtain a phase-only model by expressing the wavefunction on each site in terms of its density and phase. Substituting this into the lattice version of the driven Gross-Pitaesvkii equation, Eq. (2)
, and separating the real and imaginary parts gives 
\begin{eqnarray}
\frac{\dot{n}_k}{2n_k} = g-\Gamma n_k+J\sum_{<l>}\sqrt{\frac{n_{l}}{n_k}}\sin(\theta_{l}-\theta_k) \label{eq:dens} \\
\dot{\theta}_k = \epsilon_k + Un_k-J\sum_{<l>}\sqrt{\frac{n_{l}}{n_k}}\cos(\theta_l-\theta_k). \label{eq:phase}
\end{eqnarray} The first two terms of Eq. (\ref{eq:dens}) imply that in the absence of tunneling the occupation of each site relaxes to the steady-state value $g/\Gamma$ with rate $2g$. We consider the dynamics far enough above threshold for this to be the fastest timescale, allowing us to adiabatically eliminate the occupations by setting $\dot{n}_k=0$. We also suppose the condensates are pumped sufficiently far above threshold for the variation in occupation between sites to be relatively small, $\delta n_{kl}=n_{l}-n_{k}\ll n_k$, so that the square root factors can be replaced with one. Using these conditions to solve Eq. (\ref{eq:dens}) and substituting into Eq. (\ref{eq:phase}) gives Eq. (3) 
in the main text.

A sufficient condition for the site-to-site changes in occupations to
be small can be obtained by considering the first two terms on the
right-hand side of Eq. (\ref{eq:phase}), which balance when
$\delta n\sim (\sigma/U)$. This is an overestimate of the true
$\delta n$ because the final term, which is the kinetic energy,
spreads out the wavefunction. Thus $\delta n/n \ll 1$ certainly holds
when $(\sigma/U)/(g/\Gamma)=\sigma\alpha/g\ll 1$.

\section{Estimates of experimental parameters}

For the experiment reported in Ref. [14] 
the site-to-site variation of the energy, at the scale of a single well, is given as
$\sigma=0.03\; \mathrm{meV}$. We have estimated the tunneling $J$ by considering a pair of harmonic oscillator potentials, with minima separated by $a$, each with oscillator length $l=\sqrt{\hbar/m\omega}$. Taking $l\approx3\ \mu m$ and $a\approx 8\ \mu m$, where the synchronization transition was observed, we find $J \approx 7\ \mu \mathrm{eV}$ implying $J/\sigma \approx 0.2$. However, as the value of $J$ depends exponentially on the form of the potential there is considerable uncertainty in this estimate.

We could not reliably
estimate $\alpha=\Gamma/U=\Gamma_0/U_0$ from the data reported in [14]
, but it can be estimated from
other experiments on condensation. Ref. [32]
, using the data from [8]
, gives an estimate of $0.3$. We
obtain a similar value by relating the model in Ref.\
S\onlinecite{whittaker_coherence_2009}, and its parameters
$\gamma, n_s=25000$ and $\kappa=4\times 10^{-5}\gamma$, to the present
Gross-Pitaevskii equation. Comparing the rate equations in the two
approaches gives $\Gamma\approx \gamma/(2n_s)$, while the present
$U\approx 2\kappa$, so that we obtain $\alpha\sim 0.5$.

\subsection{}
\subsubsection{}